\documentstyle[12pt]{article}
\begin{document}

\newcommand{\ep}{\epsilon}
\newcommand{\fr}{\frac}
\newcommand{\var}{\varphi}
\newcommand{\ds}{\ant_{\Sigma}d\Sigma\,\,}
\newcommand{\reals}{\mbox{${\rm I\!\!R }$}}
\newcommand{\nats}{\mbox{${\rm I\!\!N }$}}
\newcommand{\ant}{\int\limits}
\newcommand{\intgs}{\mbox{${\rm Z\!\!Z }$}}
\newcommand{\cam}{{\cal M}}
\newcommand{\caz}{{\cal Z}}
\newcommand{\cao}{{\cal O}}
\newcommand{\cac}{{\cal C}}
\newcommand{\aaa}{\int\limits_{mR}^{\infty}dk\,\,}
\newcommand{\qpi}{(4\pi )^{\frac{q+1}2}}
\newcommand{\bbb}{\left[\left(\frac k R\right)^2-m^2\right]^{-s}}
\newcommand{\ccc}{\frac{\partial}{\partial k}}
\newcommand{\fff}{\frac{\partial}{\partial z}}
\newcommand{\iikma}{\aaa \bbb \ccc}
\newcommand{\abc}{\int\limits_m^{\infty}dk\,\,}
\newcommand{\snuu}{\sum_{n=-\infty}^{\infty}}
\newcommand{\eef}{[k^2-m^2]^{-s}}
\newcommand{\imk}{\abc\eef\ccc}
\newcommand{\ddd}{\int\limits_{mR/\nu}^{\infty}dz\,\,}
\newcommand{\eee}{\left[\left(\frac{z\nu} R\right)^2-m^2\right]^{-s}}
\newcommand{\lll}{\frac{(-1)^j}{j!}}
\newcommand{\iinma}{\ddd\eee\fff}
\newcommand{\cah}{{\cal H}}
\newcommand{\nn}{\nonumber}
\renewcommand{\theequation}{\mbox{\arabic{section}.\arabic{equation}}}
\newcommand{\komplex}{\mbox{${\rm I\!\!\!C }$}}
\newcommand{\sip}{\frac{\sin (\pi s)}{\pi}}
\newcommand{\mrnu}{\left(\frac{mR}{\nu}\right)^2}
\newcommand{\numr}{\left(\frac{\nu}{mR}\right)^2}
\newcommand{\mzs}{m^{-2s}}
\newcommand{\rzs}{R^{2s}}
\newcommand{\abl}{\partial}
\newcommand{\ablz}{{Z_D^{\nu}}'(0)}
\newcommand{\g}{\Gamma\left(}
\newcommand{\zzz}{\int\limits_{\gamma}\frac{dk}{2\pi i}\,\,}
\newcommand{\yyy}{(k^2+m^2)^{-s}\frac{\partial}{\partial k}}
\newcommand{\ikma}{\zzz \yyy}
\newcommand{\ead}{e_{\alpha}(D)}
\newcommand{\sul}{\sum_{l=-\infty}^{\infty}}
\newcommand{\tint}{\int\limits_0^{\infty}dt\,\,}
\newcommand{\nsum}{\sum_N}
\newcommand{\sual}{\sum_{\alpha =1}^{D-2}}
\newcommand{\sulnu}{\sum_{l=0}^{\infty}}
\newcommand{\sujnu}{\sum_{j=0}^{\infty}}
\newcommand{\suani}{\sum_{a=0}^i}
\newcommand{\suanzi}{\sum_{a=0}^{2i}}
\newcommand{\zend}{\zeta_D^{\nu}}
\newcommand{\amed}{A_{-1}^{\nu ,D}(s)}
\renewcommand{\and}{A_{0}^{\nu ,D}(s)}
\newcommand{\pol}{\left(\frac 1 {s+1/2}+\ln R^2\right)}
\newcommand{\nen}{1+e^{2\pi\nu}}
\newcommand{\inu}{\int\limits_0^{\infty}d\nu\,\,}
\newcommand{\nr}{\nu^2-m^2R^2}
\newcommand{\aid}{A_{i}^{\nu ,D}(s)}
\def\beq{\begin{equation}}
\def\eeq{\end{equation}}
\def\bea{\begin{eqnarray}}
\def\eea{\end{eqnarray}}

\begin{titlepage}

\title{
{\Large \bf Casimir energy of a massive field in a genus-1 surface}}
\author{
K. Kirsten\thanks{E-mail address: kirsten@tph100.physik.uni-leipzig.de}\\
Universit{\"a}t Leipzig, Institut f{\"u}r Theoretische Physik,\\
Augustusplatz 10, D-04109 Leipzig, Germany\\
\\
and\\
\\
E. Elizalde\thanks{Permanent address:
Center for Advanced Study CEAB, CSIC, Cam\'{\i} de Santa
B\`arbara, 17300 Blanes,
Spain;
e-mail: eli@ecm.ub.es} \\
Dipartimento di Fisica, Universit\`a degli Studi di Trento, \\
I-38050 Povo, Trento, Italia \\
 }

\thispagestyle{empty}

\vspace*{-1mm}

\maketitle

\vspace*{1mm}

\begin{abstract}
We review the definition of the Casimir energy steming naturally
from
the concept of functional determinant through the zeta function prescription.
This is done by considering the theory at finite temperature and by
defining then the Casimir energy as its energy in the limit $T\to 0$. The
ambiguity in the coefficient $C_{d/2}$ is understood to be a result of the
necessary renormalization of the free energy of the system. Then,
as an exact, explicit example never calculated before,
the Casimir energy for a massive scalar field living in
a general $(1+2)$-dimensional toroidal spacetime
                           (i.e., a general surface of genus one)
with flat spatial geometry
---parametrized by the corresponding Teichm\"uller parameters--- and
  its precise
dependence on these parameters and on the
mass of the field is obtained under the form of an
analytic function.

\end{abstract}
\end{titlepage}

\newpage

\section{Introduction}
\setcounter{equation}{0}
\setcounter{page}{2}
The Casimir effect \cite{casimir48} is a beautiful and simple manifestation
of the influence that boundaries or non-trivial spacetime topologies have
on quantum field theories (see for example
\cite{plunienmullergreiner86,ambjornwolfram83,eorbz,moste}). The modern
approach for the calculation of the Casimir energy is the zeta function
regularization scheme put forward in \cite{blauvisserwipf88}. The idea is
the following. In order to have a well-defined notion
of energy, let us work in a $d$-dimensional ultrastatic spacetime
\cite{dewitt75}. Thus the metric in some coordinate system is of the form
$g=-(dx^0)^2 +g_S$ with the spatial part $g_S$ of the metric (see
Eq.~(\ref{eq:2.1})).
The differential operator $D$ describing the field
equation may be decomposed as $D=-\partial _0^2 +D_S$. Introducing
$E_n^2=\lambda_n$ as the eigenfrequencies of $D_S$, the zero-point energy
is formally given by
\beq
E_{Casimir} =\frac 1 2 \sum_n E_n. \label{eq:1.1}
\eeq
It may be regularized by defining
\beq
E_{reg}(\epsilon ) =\frac 1 2 \mu^{2\epsilon}
             \zeta _S (-1/2 +\epsilon ) , \label{eq:1.2}
\eeq
with $\zeta _S (s)$ being the zeta function associated with the
($d-1$)-dimensional operator $D_S$. The scale $\mu$ with dimension
(length)$^{-1}$ has to be introduced in order to keep the zeta function
dimensionless for all $s$.

General zeta function theory \cite{voros87} tells us, that $E_{reg}
 (\epsilon ) $ is a meromorphic function with a pole at $\epsilon =0$,
its residue being $-(1/2) C_{d/2} (D_S)/(4\pi )^{d/2}$. The coefficient
$C_{d/2}
(D_S)$ is the Seeley-De Witt coefficient appearing in the asymptotic
expansion for small $t$ of the heat-kernel associated with $D_S$,
\bea
K(t) &=& \sum_n  e^{-\lambda_n t}\nn\\
   &\sim & \left(\frac 1 {4\pi t}\right)^{(d-1)/2} \sum_{l=0,1/2,1,...}
^{\infty}C_l(D_S) t^l .\label{eq:1.3}
\eea
The pole appearing in Eq. (\ref{eq:1.2}) has to be absorbed into the bare
action which thus must contain a term proportional to $C_{d/2} (D_S)$.
It is clear then, that the Casimir energy has an ambiguity proportional to
$C_{d/2} (D_S)$ (we will come back to this point later).

Adopting the minimal subtraction scheme, one defines
\bea
E_{Casimir} &=& \frac 1 2 \lim_{\epsilon \to 0}
    \frac 1 2 \mu^{2\epsilon}\left[\zeta_S (-1/2 +\epsilon )
+\zeta_S (-1/2-\epsilon )\right]\nn\\
&=& \frac 1 2 \left[ PP\,\, \zeta_S (-1/2)-\frac{C_{d/2}(D_S)
             }{(4\pi)^{d/2}}\ln\mu^2\right] , \label{eq:1.4}
\eea
where the symbol $PP$ stands for taking the principal part. Based on this
definition, during the last years the Casimir energy has been calculated for
a  variety of examples
\cite{eorbz,dolannash92,carusonetosvaitersvaiter91}.

The definition (\ref{eq:1.4}) of the Casimir energy is completely
equivalent to the one steming naturally from
the definition of functional determinant by the zeta function
prescription (for very recent
considerations on this issue see \cite{zds}).
This may be done by considering the theory at finite temperature and by
defining the Casimir energy as its energy in the limit $T\to 0$,
this idea going back to Gibbons \cite{gibbons77}
who considered the single quantum mechanical oscillator. In the more general
context of quantum field theory under some external conditions like boundaries
or gravitational fields this definition has been employed for example in
\cite{dowkerkennedy78,ambjornwolfram83,dowker84,cognolavanzozerbini92,trento}).
The
ambiguity in the coefficient $C_{d/2}$ may be understood to be a result of the
necessary renormalization of the free energy of the system.

Having summarized the main arguments in favor of the definition of the Casimir
energy as given by Eq.~(\ref{eq:1.4}), in section 3 we will present the
calculation of the Casimir energy
for a massive scalar field in a general $(1+2)$-dimensional
toroidal spacetime with flat spatial geometry. The general flat geometry will
be
parametrized by the corresponding two Teichm\"uller parameters and the complete
dependence of the Casimir energy on these parameters
and on the mass of the field will be obtained under the form of an analytic
function, by using the extended
Chowla-Selberg zeta function formula derived
by one of us in Refs.~\cite{eecs1,eecs2}.
For the massless case we obtain complete agreement with previous results by
Dowker \cite{dowker89a}.

\section{Zeta function definition of the Casimir energy}
\setcounter{equation}{0}
Let us first briefly
summarize the motivation for the definition (\ref{eq:1.4}) of
the Casimir energy.
For definiteness, let us consider the quantum field theory of a free
scalar field in curved spacetime, the Dirac field may be treated analogously.
As mentioned before, in order to have
a well defined notion of energy
we shall restrict our considerations
to a $d$-dimensional ultrastatic spacetime ${\cal M}$, possibly with a
boundary,
and with the metric
\begin{eqnarray}
ds^2=-dx_0^2+g_{ab}(\vec x )dx^adx^b,\label{eq:2.1}
\end{eqnarray}
where
$\vec x =(x_1,...,x_{d-1})$.  The action of the field theory we consider is
\cite{dewitt75}, \cite{critchleydowkerkennedy80},
\begin{eqnarray}
S=-\frac 1 2 \ant_{\cam} d^d x |g|^{\frac 1 2}\phi^{\dagger}(x)
\left(\Box -\xi R-m^2\right)\phi (x),\label{22}
\end{eqnarray}
with  $\Box$ being  the Laplace-Beltrami
operator of the ultrastatic spacetime.  Variation of equation
(\ref{22}) subject to the constraints
\begin{eqnarray}
\delta \phi (x')&=&0,\nn\\ n^{\mu '}\nabla _{\mu '}\delta \phi
(x')&=&0,\nn
\end{eqnarray}
where the prime refers to quantities
defined on the boundary $\partial {\cal M}$, yields the equation of
motion
\begin{eqnarray}
\left(\Box -\xi R-m^2\right) \phi
(x)=0,\label{25}
\end{eqnarray}
which is the generalized Klein-Gordon equation.
The following discussion will be quite
general, so the boundary condition need not to be specified at this
point. A unique boundary value problem is posed, for example, by
assuming Dirichlet- or Robin-boundary conditions on the field.

In an ultrastatic spacetime the quantum field theory
at finite temperature
may be developed in
complete analogy with the Minkowski-space theory and that
is why we will skip the
 details of the calculation.
In the Euclidean formulation of the finite temperature theory, the partition
sum
$\caz$
may be written under the form of a functional integral of the kind
\beq
{\cal Z} [\beta ] =\int [d\var]\exp\left\{-\frac 1 2 (\var ,D\var
)\right\},
\eeq
where we have used
the scalar product
\beq
(f,h) =\ant_0^{\tau} d\tau \ds |g|^{1/2} f^* h\nn
\eeq
for the two vectors $f$ and $h$, and being $\Sigma$  the spatial section of the
manifold
$\cam$.
Here the integration extends over all fields periodic in the imaginary
time $\tau$ with periodicity $\beta =1/T$ and fulfilling the boundary
conditions at the spatial boundary. The operator $D$ is given by
\beq
D =-\frac{\partial^2}{\partial \tau^2} -\Delta +\xi R +m^2.
\eeq
Then, one formally obtains
\beq
{\cal Z} [\beta , \mu
] =(\det \lambda ^{-2} D)^{-\frac 1 2},
\eeq
 the scale $\lambda$ \cite{hawking77} being  necessary in order to keep
everything
dimensionless. The functional determinant of the operator
$\lambda^{-2} D$ needs, of course, regularization.
We will use the zeta-function regularization scheme introduced
by Dowker, Critchley \cite{critchleydowker76} and Hawking
\cite{hawking77}.
In this scheme, the free energy is defined as
\beq
F[\beta ]=-\frac 1 {2\beta} \left[\zeta_d (0,\beta )\ln \lambda^2
+\zeta_d ' (0,\beta )\right].\label{free}
\eeq
The function $\zeta_d(s,\beta)$ is the zeta-function
associated with the operator
$D$. Using the ansatz
\begin{eqnarray}
u_{l,k}=\frac 1 {\beta} \exp\left(\frac{2\pi
il}{\beta}\right)g_k(\vec x)\nonumber
\end{eqnarray}
the eigenvalues are seen to have the form
\begin{eqnarray}
\nu_{l,k}^{\pm}=\left(\frac{2\pi
l}{\beta}\right)^2+E_k^2,\,\, \ \ l\in{\intgs},\nonumber
\end{eqnarray}
with the energy eigenvalues being defined through
\beq
(-\Delta +\xi R +m^2) \psi_k (\vec x )=E_k^2 \psi_k (\vec x).
\eeq
Making use of a Mellin-transformation and a theta-function identity
\cite{hille62}, the free energy may be written in the form
\begin{eqnarray}
F[\beta]&=&\frac 1 2 PP\zeta_S\left(-\frac
1 2\right)+\frac 1 {2(4\pi)^{\frac d 2}}C_{\frac d 2}[\ln
\lambda^2-1+2\ln 2]\label{endre}\\ &+&\frac 1 {\beta}
\sum_j\ln\left(1-e^{-\beta E_j}\right)\nn.
\end{eqnarray}
The energy of the system is then given by
\begin{eqnarray}
E=\frac{\partial}{\partial\beta}\beta F[\beta ].
\label{energie} \end{eqnarray}
Finally, defining the Casimir-energy as the limit of
(\ref{energie}) for $T\to 0$, we obtain
\begin{eqnarray}
E_{Cas}[\partial{\cal M}]=\lim_{T\to 0}E=\frac 1 2
 PP\zeta_S\left(-\frac 1
2\right)+\frac 1 {2(4\pi)^{\frac d 2}}C_{\frac d 2}\ln
\tilde{\lambda}^2,\label{casimir}
\end{eqnarray}
with $\tilde{\lambda}=\frac{2\lambda}{\sqrt{e}}$. We thus arrive to
the definition
of Refs.~\cite{dowker84,blauvisserwipf88}.
In the last reference a detailed
discussion
of the meaning of this definition and of
the problem of renormalization has
been
carried out.
 From the above derivation of the definition of the Casimir energy
it is completely clear that the ambiguity of the Casimir energy is
simply a result of the (in general) necessary renormalization
of the free
energy. In the case when the coefficient $C_{d/2}$  vanishes,
the definition (\ref{casimir}) gives a well defined, finite value.
\section{Casimir energy in a (1+2)-dimensional toroidal spacetime}
\setcounter{equation}{0}
Let us consider, as an example, a (1+2)-dimensional spacetime with the
topology $\reals \times
T^2$ \cite{seriu}. We will concentrate on the case when the geometry of
the space
$\Sigma\simeq T^2$ is locally flat. One can construct such  geometry,  as is
usual, by taking $\Sigma =[0,1]\times [0,1]/\sim $, where the equivalence
relation is defined by $(\xi_1 ,0) \sim (\xi_1 ,1)$ and $ (0,\xi _2 )\sim
(1,\xi_2)$. A flat $2$-geometry is endowed on $\Sigma$ by giving  it a metric
\beq
ds^2 =h_{ab} d\xi^a d\xi^b,
\eeq
where
\beq
h_{ab}=\frac 1 {\tau_2}\left(
   \begin{array}{cc}
     1 & \tau_1\\
      \tau_1 & |\tau|^2
   \end{array}\right)  .
\eeq
The
$(\tau_1, \tau_2)$ are the Teichm\"uller parameters, independent of the spatial
coordinates
$(\xi_1,\xi_2)$,
and $\tau =\tau_1+i\tau_2$, $\tau_2>0$ \cite{hatfield92,luesttheisen89}.

The Laplace-Beltrami operator of this metric is given by
\beq
\Delta =-\frac 1 {\tau_2} (|\tau|^2 \partial_1^2 -2\tau_1\partial_1\partial_2
+\partial_2^2 ),
\eeq
being its eigenvalues
\beq
\lambda_{n_1,n_2} =\frac{4\pi^2}{\tau_2} (|\tau|^2n_1^2 -2\tau_1n_1n_2
   +n_2^2 ).
\eeq
In the  massive case, $m \neq 0$ the spectrum runs over $n_1,n_2 \in \intgs$.
In  the massless case the zero-mode of $\Delta$, $n_1=n_2=0$,
has to be
excluded.

An exact analysis of the Casimir energy for this spacetime is possible since it
reduces to
a case of the Chowla-Selberg zeta function (when $m=0$) or to one of the
extended formula
that has been obtained recently (case $m\neq 0$).

In fact, in the case when $m \neq 0$, the relevant formula is a particular
application of the following. Let us consider the double series
\beq
E(s;a,b,c;q) \equiv
{\sum_{m,n \in \mbox{\bf Z}}}' (am^2+bmn+cn^2+q)^{-s}, \label{1}
\eeq
 with $q\neq 0$ (in general),
the parenthesis in  (\ref{1}) is the
inhomogeneous quadratic form
\begin{equation}
Q(x,y)+q, \ \ \ \  Q(x,y)  \equiv ax^2+bxy+cy^2,
\end{equation}
restricted to the  integers. In the general theory that deals
with the homogeneous case, one assumes that $a,c >0$ and that
the discriminant
\beq
\Delta =4ac-b^2 >0
\eeq
(see \cite{cs}).
Here we will impose the additional condition that $q$ be such that
 $Q(m,n)+q \neq 0$, for all $ m,n \in
$ {\bf Z}. In the usual applications of the theory, those
conditions are indeed satisfied.
For the analytical continuation of (\ref{1}),
the following expression has been obtained in Refs. \cite{eecs1,eecs2}
 \beq
 E(s;a,b,c;q) =
  2\zeta_{EH} (s,q/a)\, a^{-s} + \frac{2^{2s}
\sqrt{\pi}\, a^{s-1}}{\Gamma (s) \Delta^{s-1/2}} \, \Gamma (s -
1/2) \zeta_{EH} (s-1/2,4aq/\Delta)
\ \ \ \ \ \ \ \ \label{gcs} \eeq
\[ +
\frac{2^{s+5/2} \pi^s }{\Gamma (s) \sqrt{a}}
\sum_{n=1}^\infty
n^{s-1/2} \cos (n \pi b/a) \sum_{d|n} d^{1-2s}
 \left( \Delta + \frac{4aq}{d^2} \right)^{-s/2+1/4}
K_{s - 1/2}\left( \frac{\pi n}{a}
\sqrt{ \Delta + \frac{4aq}{d^2}} \right),
\]
 $\sum_{d|n}d^s$
denoting the sum over the divisors of $n$ and
where the function $\zeta_{EH} (s,p) $ (the one dimensional
Epstein-Hurwitz or inhomogeneous Epstein series) is given by
 \begin{eqnarray}
 \zeta_{EH}(s;p) &=& \sum_{n=1}^\infty  \left( n^2 + p
\right)^{-s} = \frac{1}{2} {\sum_{n \in \mbox{\bf Z}}}' \,
 \left( n^2 + p
\right)^{-s} \label{zeh1} \\
& =& -\frac{p^{-s}}{2} + \frac{\sqrt{\pi} \, \Gamma (s-
1/2)}{2\, \Gamma (s)}
p^{-s+1/2} + \frac{2\pi^s p^{-s/2 +1/4}}{\Gamma (s)}
\sum_{n=1}^\infty n^{s -1/2} K_{s -1/2} (2\pi n\sqrt{p}).
\nonumber
\end{eqnarray}
Eq. (\ref{gcs}) provides the analytical continuation of the
inhomogenous Epstein series, in the variable $s$, as a meromorphic
function in the complex plane. Its pole structure is explicitly given
in terms of the well-known pole structure of $\zeta_{EH} (s,p)$.

Eq. (\ref{gcs}) has been found by one of us and called the {\it extended
Chowla-Selberg} formula, since it contains the Chowla-Selberg formula
 as the particular case $q=0$,
i.e.,
\begin{eqnarray}
&& E(s;a,b,c;0) = 2\zeta (2s)\, a^{-s} + \frac{2^{2s}
\sqrt{\pi}\, a^{s-1}}{\Gamma (s) \Delta^{s-1/2}} \,\Gamma (s
-1/2) \zeta (2s-1) + \frac{2^{s+5/2} \pi^s }{\Gamma (s)
\Delta^{s/2-1/4}\sqrt{a}}
\nn \\ && \hspace{1cm} \times \sum_{n=1}^\infty n^{s-1/2} \, \sigma_{1-2s} (n)
\,
 \cos (n \pi b/a)  \, K_{s - 1/2}\left( \frac{\pi n \sqrt{\Delta}}{a} \right).
\label{cs1}
\end{eqnarray}
where
\begin{equation}
\sigma_s(n) \equiv \sum_{d|n} d^s.
\end{equation}
Formula (\ref{gcs}) has been obtained for the first time in \cite{eecs1}, and a
misprint has been corrected in \cite{eecs2}.
The good convergence properties of expression (\ref{cs1}),
that were so much
prised by Chowla and Selberg, are shared by its
non-trivial extension (\ref{gcs}). This renders the use of the
formula quite
simple. In fact, the two first terms are just nice
---under the form  (\ref{zeh1})--- while
the last one (impressive in appearence) is even more  quickly
convergent than in the case of Eq. (\ref{cs1}),
and thus absolutely harmless in practice. Only a few
first terms of the three series of Bessel functions in (\ref{gcs}),
 (\ref{zeh1}) need to be
calculated, even if one demands good accuracy. We
should also notice that the only pole of (\ref{cs1})
 at $s=1$ appears through
$\zeta (2s-1)$ in the second term, while for $s=1/2$, the
apparent singularities of the first and second terms cancel each
other and no pole is formed. Analogously, the pole at $s=1/2$ in
(\ref{gcs}) comes only from the first term.
Eq. (\ref{gcs}) also has these good
properties, {\it for any non-negative value of} $q$.
In fact, for
large $q$ the convergence properties of the series of Bessel functions
are clearly enhanced, while for $q$ small we get back to the case of
Chowla and Selberg. Notice, however, that this is not obtained through
the high-$q$ expansion (e.g., just putting $q=0$ in (\ref{gcs})), but
 using a low-$q$, binomial expansion of the kind
\beq
\sum_{n=0}^{\infty} \left[ a(n+c)^2+q\right]^{-s}= a^{-s}
\sum_{m=0}^{\infty} \frac{(-1)^m \Gamma(m+s)}{\Gamma (s) \, m!}
\left( \frac{q}{a} \right)^m \zeta_H (2s+2m,c),
\eeq
which is convergent for $q/a \leq 1$. For $q \rightarrow 0$ it
reduces to  $a^{-s} \zeta_H (2s,c)$.
 Actually, formula (\ref{gcs})
 is still valid in a domain of
negative $q$'s, namely for $q > -\min (a,c,a-b+c)$.

Turning now to the particular application of the formula in our specific
situation,
we see that the zeta function corresponding to the Laplace-Beltrami operator in
the
massive (resp. massless) case is simply given by:
\beq
\zeta_{\Delta + m^2} (s) = m^{-2s} + \left( \frac{4\pi^2}{\tau_2} \right)^{-s}
E\left(s; |\tau|^2, -2\tau_1,1; \frac{\tau_2m^2}{4\pi^2} \right)
\eeq
and
\beq
\zeta_{\Delta} (s) =  \left( \frac{4\pi^2}{\tau_2} \right)^{-s}
E\left(s; |\tau|^2, -2\tau_1,1; 0 \right),
\eeq
respectively.
The values at $s=-1/2$ are finite and define the corresponding Casimir
energy.
After
performing the necessary calculations, and according to the prescription
that has
been derived in the first part of this paper, we obtain as result the
quite simple
expressions
\bea
&& \zeta_{\Delta + m^2} (-1/2) = -\frac{m^3}{6\pi} - \frac{2m}{\pi}
\sum_{n=1}^\infty n^{-1} K_1 (nmx_2) - \sqrt{2} \left( \frac{m
x_2}{\pi}\right)^{3/2}
\sum_{n=1}^\infty n^{-3/2} K_{3/2} (nm/x_2) \nn \\
&& -8 x_2\sum_{n=1}^\infty n^{-1} \cos (2\pi n x_1^2)
 \sum_{d|n} d^{2}
 \sqrt{ 1 + \frac{m^2}{(2\pi x_2 d)^2}} \ K_1 \left( 2\pi n x_2^2
 \sqrt{ 1 + \frac{m^2}{(2\pi x_2 d)^2}} \right), \label{d12m}
\eea
and
\beq
 \zeta_{\Delta } (-1/2) = -\frac{\pi}{3x_2} + 4\pi \zeta'(-2) x_2^3
 -8 x_2\sum_{n=1}^\infty n^{-1} \cos (2\pi n x_1^2) \sigma_2(n) \,
 K_1 \left( 2\pi n x_2^2 \right), \label{d12}
\eeq
in terms of the variables
\beq
x_1 = \sqrt{\frac{\tau_1}{\tau_1^2 + \tau_2^2}}, \qquad
x_2 = \sqrt{\frac{\tau_2}{\tau_1^2 + \tau_2^2}}. \label{vch}
\eeq

The extrema of the corresponding Casimir energy for the case $m^2=100$,
 in terms of the original
Teichm\"uller coefficients $\tau_1$ and $\tau_2$, are to be read from Figs.
1-3.
 In the three-dimensional plot over the plane
$\tau_1,  \tau_2$ (Fig. 1),  the maximal Casimir energy is seen to be localized
on the section
 $\tau_2=1$. In order to show this fact more clearly, in Fig. 2 we have
represented the section
 $\tau_1=0$, but the situation is common to any section
$\tau_1=$ const. On the section  $\tau_2=1$  a periodic structure appears
asociated with the value of
 $\tau_1$ along this section (Fig. 3). This behavior is easy to recognize from
the form of
 the function
$\zeta_\Delta (-1/2)$, (\ref{d12}), and is common to {\it any} section
$\tau_2=$ const.
All the figures depicted here have been
 obtained taking the first 20 terms from the sum over $n$ in (\ref{d12}).

 \vspace{5mm}


\noindent{\large \bf Acknowledgments}

We thank all the members of the Department of Theoretical Physics of the
 University of Trento for
warm hospitality,  specially Sergio Zerbini, Luciano Vanzo, Guido Cognola,
Ruggero Ferrari
 and Marco Toller.
Thanks are given to the referee of a previos version of this paper
for very precise comments that led to its improvement.
This work was
finished while  EE was visiting the Institute of Theoretical Physics,
Chalmers University of Technology (Sweden), and has been
supported by DGICYT and Ministry of Foreign Affairs
(Spain), by CIRIT (Generalitat de Catalunya) and by INFN
(Italy).

\newpage


\newpage

\noindent{\large \bf Figure captions}
\vspace{6mm}

\noindent{ \bf Fig. 1}

 Three-dimensional plot over the plane
$\tau_1,  \tau_2$, of the Casimir energy  for the case $m=10$.
 One clearly observes that the extremal Casimir energy is
 localized on the section $\tau_2=1$. All the figures have been
 obtained taking the first 20 terms from the sum over $n$ in (\ref{d12}).

\bigskip

\noindent{ \bf Fig. 2}

Plot of the section $\tau_1=0$, where the fact that the extremal Casimir energy
is obtained for
 $\tau_2=1$ is seen undoubtedly.  This behavior is common to  any section
$\tau_1=$ const.

\bigskip

\noindent{ \bf Fig. 3}

Plot of the section $\tau_2=1$. Here a  periodic structure appears asociated
with the value of
 $\tau_1$ along this section.  This behavior is common to any section $\tau_2=$
const.

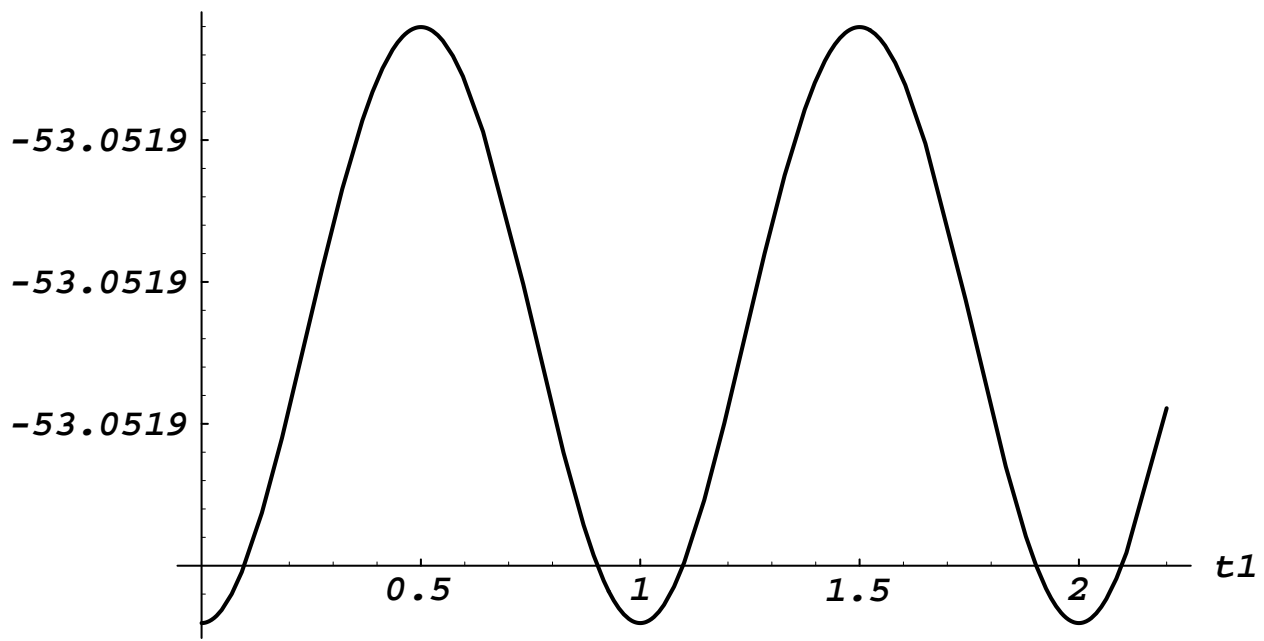
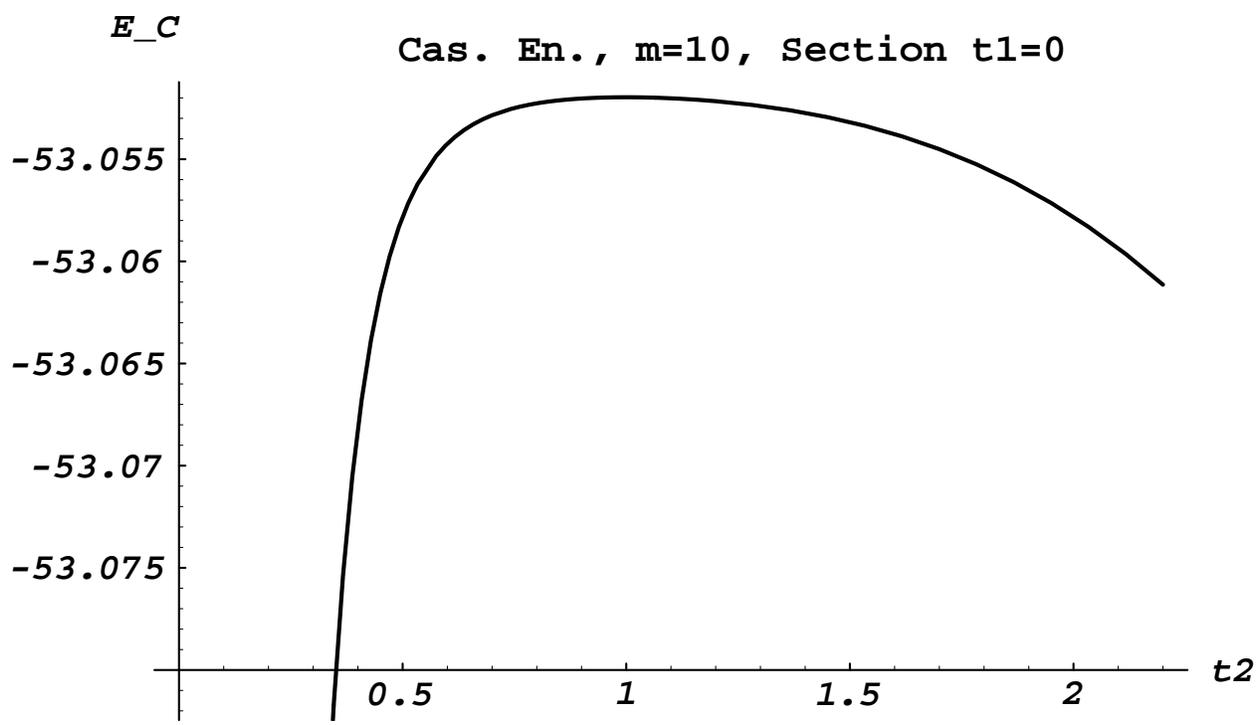
\end{document}